# What surfaces in the operation of noble liquids dark matter detectors

**S. Pereverzev**

*Lawrence Livermore National Laboratory,*
*7000 east Ave, Livermore, California, USA.*
*E-mail*: pereverzev1@llnl.gov

ABSTRACT: Though noble element dual-phase detectors have a long application history in dark matter searches, some uncertainties and differences in backgrounds persist. We compare effects caused by unextracted electrons on the liquid-gas interface in Xe and Ar dual-phase detectors with a large family of phenomena at the liquid helium surface. We pose that electron and ion accumulation on the liquid surface in detectors can lead to the formation of ordered surface states, charged liquid surface instabilities in an electric field, electrospraying, interactions with surface waves, and other effects. Not only delayed electron emission signals can be generated, but the extraction efficiency for electrons produced below the liquid surface can be altered by the presence of surface charges. Several factors lead to surface electron accumulation, and problems can become more severe with the increased detector size. We discuss possible experiments to reveal surface electron effects and design changes to alleviate electron accumulation. We conclude that studies of these effects are desirable before making final design decisions for the new multi-ton liquid Xe dark matter detector projects like DARWIN, XLZD, and large Ar dual-phase detectors.

# Contents

## 1. Introduction

Over the last 20 years, the community has continuously increased the liquid mass of dual-phase dark matter particle detectors from 10 kg in the Xenon10 detector [1] to about 7 tons in the LZ experiment [2]. Building experiments with up to 30-100 tons of liquid Xenon targets- like the DARWIN [3] project- are now under discussion.

Fast and unimpeded transport of electrons and photons without diverting energy for chemical reactions, long-living excitations, trapping, and delayed releases of energy and charges allows multiple detector applications of noble elements. In dual-phase detectors, photon detector arrays detect the scintillation pulse S1 produced in liquid by the energetic particle. Free electrons produced by the primary particle move in the applied electric field to the liquid surface, escape into gas and generate the electroluminescence pulse S2. Each electron extracted into gas produces multiple photons (gas amplification), allowing single-electron detection and better energy resolution (electron number resolution) than measurements of small current pulses. Low energy detection threshold, electron and nuclear recoil discrimination by the ratio of S1/S2 pulses, and scalability to sizeable liquid target mass make the technology of choice for direct dark matter particle searches.

At the same time, the appearance of significantly delayed multiple electron emission events (e-bursts) after muons and other large ionization events in some detectors like Xe10 [1] and LUX [4], the origin of the excess in few-electron events (1,2,3,…,8) [5], and differences in observed low-energy events spectra, are not well understood.

We start with comparing the accumulation of unextracted electrons on the liquid-gas interface in different devices: differences we see and factors that impede the escape of electrons and ions trapped at the liquid surface to the detector walls.

We briefly mention the effects known for liquid helium surface: formation of ordered electron states like Wigner crystal and multi-electron dimple lattice, charged liquid instabilities in a strong electric field, electrospraying-like effects, and interactions with surface waves. The appearance of ordered surface states can explain differences in unextracted electron behavior in detectors and why electron bursts in some experiments can occur at the exact position of the previous S2 event with a significant time delay. We also discuss the possibility that accumulating unextracted electrons on the surface can suppress the extraction of small electron signals originating below the liquid surface.

More features of the electron extraction process and delayed electron emission in detectors have no clear explanation yet. The effects discussed in this paper can lead to uncertainties and even misinterpretations or wrong conclusions about particle physics in large detector experiments. On the other hand, one can detect and study these phenomena in condensed-matter-style experiments with a small detector set-up. We believe that through systematic studies of material and condensed-matter effects, we will better understand the detector's operation and resolve the problem with annoying excess backgrounds in noble liquid [5] and solid-state [6] dark matter particle detectors.

**2. Dwelling time of unextracted electrons on the liquid surface**

An electron can be attracted to dielectric fluid due to fluid polarization. The kinetic energy of a free electron moving in a liquid under an applied electric field must be above ~0.85 eV for the electron to escape from liquid Xe and above ~0.65 eV for the liquid Ar; see, for example, [7,8]. Free electron drift in noble liquids like Ar, Kr, and Xe under an applied electric field can be described as chaotic motion with multiple direction changes in collision events [7,8]. If a field-driven electron loses kinetic energy near the liquid surface, it will not escape and stay on the liquid-gas boundary.

We must distinguish between free surface electrons, presumably having large surface mobility, and negative atomic and molecular ions, which can also be trapped on the liquid surface but will have low surface mobility. When free surface electrons stay at the surface sufficiently long, they can be trapped by electronegative impurities and form low-mobility negative ions. Unfortunately, we do not know publications where surface mobilities of free surface electrons and surface ions were determined, nor any data on the lifetime of the free surface electron before trapping by electronegative impurities.

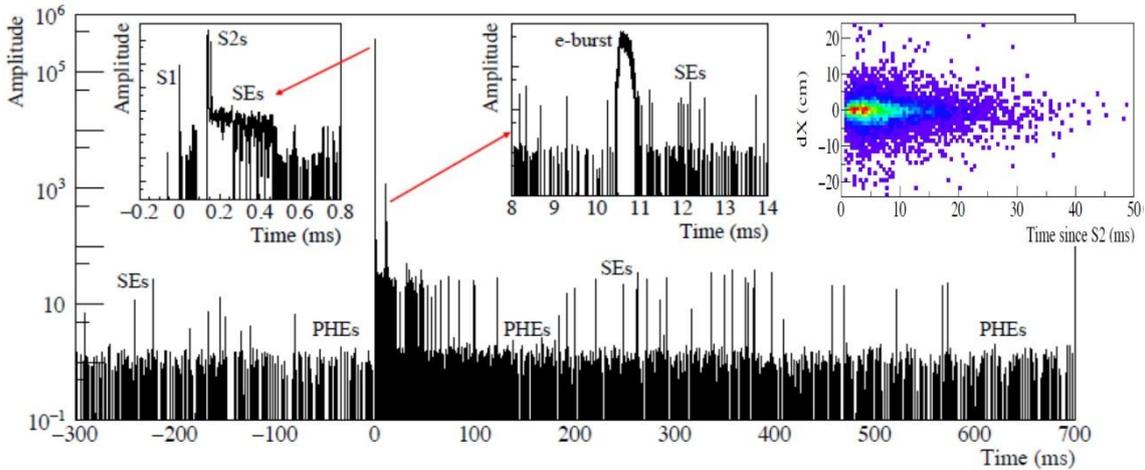

**Figure 1.** A continuous LUX waveform over one second (from [4]). With only one 2.3 MeV gamma interacts with the detector, thousands delayed background electrons and photons were observed following S1 and S2 pulses. Left and central inserts show zoomed S1-S2 event window and delayed e-burst events and random single electrons (SE) and photoelectrons (PHE) pulses. The right insert is the X position difference between e-bursts and preceding S2 as a function of time delay (only events with no additional S2 pulses during [-30,+50] ms relative to S2 were taken)

Negative ions cannot gain considerable kinetic energy when drifting in an electric field in liquid. Their short-range interaction with neutral noble element atoms (Xe, Ar) can be repulsive [9]. We cannot exclude the possibility of a lower energy barrier for the extraction of negative ions into a gas (lower than the barrier for free electron extraction) in the presence of an extraction field and a high surface concentration of negative surface charges.

Electrons and negative ions trapped at the liquid-gas interphase can drift away to the detector's metal walls or be extracted into the gas by some processes.

One type of process - e-bursts- is a delayed escape event of a significant number of electrons. It may include the escape of negative ions with subsequent liberation of electrons in ion-gas atoms collisions. E-bursts were observed in Xenon 10 [1] and later in other detectors and were studied in detail in the LUX detector [4]. Notably, in the LUX detector, e-burst appeared at the exact X-Y coordinates where previous significant S2 events (electron extraction) took place with delays up to 50 ms (see Fig.1); several e-burst were possible at the same location [4]. Thus, unextracted electrons can stay at the liquid surface for up to 50 ms (or more, no longer data were analyzed) in LUX without drifting from the "origin" X-Y position.

The RED1 (Russian Emission Detector) [10], ZEPLIN III [11], and RED100 [12] detectors have no E-bursts. In these detectors, the anode grid in gas is attached to a metal ring holder which is touching the liquid surface, see Fig.2. The calculated electric field pattern in the ZEPLIN III detector [11] has an electric field component tangential to the liquid surface near the active area's perimeter (see fig.3), This field component sweeps surface electrons or ions nearing the perimeter area toward the metal anode support ring. While E-bursts are not present in these detectors, the paper [6] describes intense light emission pulses (S3 pulses) at the perimeter of the RED1 detector following muons or other significant ionization events.

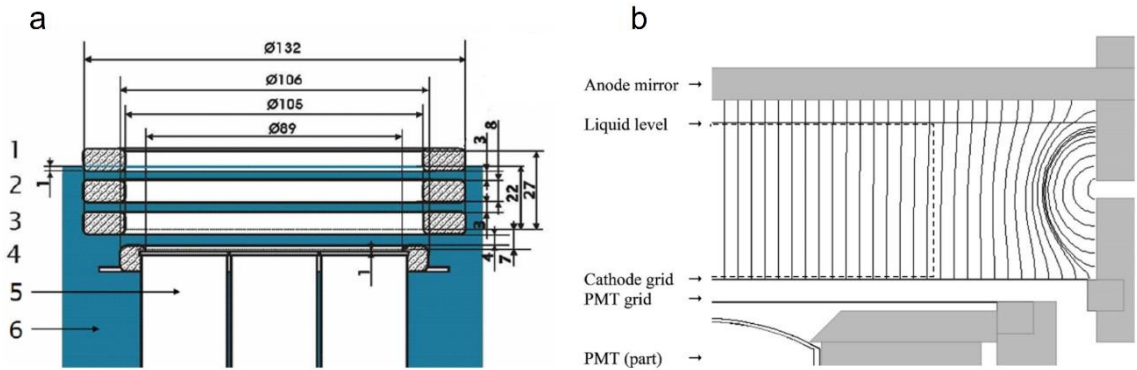

**Figure 2.** Left-electrode system in RED1 detector [10]: 1-first electrode ring and aluminum-coated stainless steel mirror on it; 2-intermediate field-shaping ring; 3-third electrode ring with mesh cathode (stainless steel); 4- electrode ring with screening grid (stainless still); 5-PMTs; 6- LIQUID Xe; all dimensions in mm.
Right- enlarged fragment of ZEPLINIII design [11] (RED1 was a prototype for ZEPLIN III) with calculated electron trajectories close to the field-shaping electrodes; the dashed line shows the boundary of the fiducial volume; at the perimeter of the detector field component tangential to the liquid surface is pushing surface charges toward the metal electrode.

For muon events in the RED1 detector [8], the authors separated the time between S2 and S3 events into shorter intervals and calculated the position of light production (center of gravity) for each interval. The light production was moving along a straight line connecting the S2 and S3 locations, indicating the presence of some electron emission mechanism different from e-bursts or S3 events. S3 pulses were present when the liquid level in the detector was slightly below the anode support ring [13]. So, the S3 light pulses originated in the strong electric field region in a thin gas layer between the charged liquid surface and the metal anode support ring. As shown in Fig. 2, the grid holder design was slightly changed in the ZEPLIN III detector relative to the RED1 detector– likely to avoid accidental separation of the anode holder ring from the liquid surface.

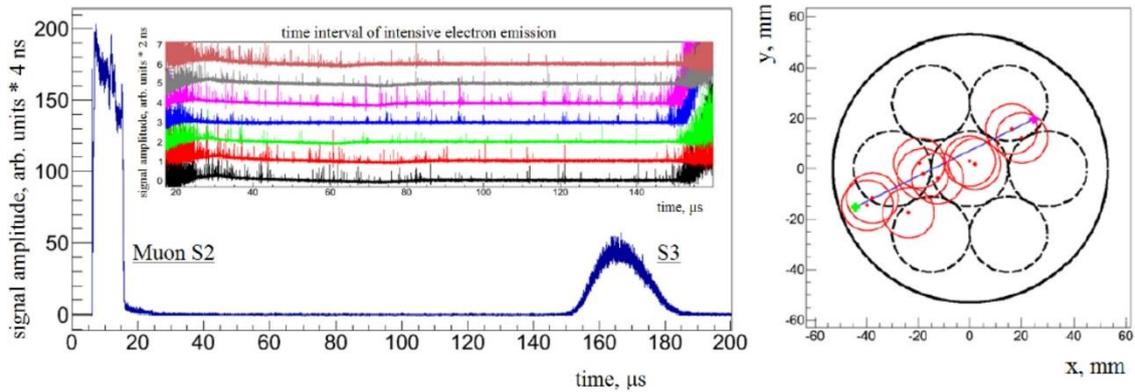

**Figure 3.** Left- Example of events containing a muon S2 signal and the following S3, the sum of waveforms in all seven PMTs in RED1 [10]. On insert - individual PMT waveforms between S2 and S3 events with electron emission signals during this period; S2 is essentially distorted due to saturation. Right- reconstruction of the spatiotemporal image of the muon event, projection onto a horizontal plane. The Magenta point is muon's effective position; red dots correspond to subinterval position of light production position (red circles represent position reconstruction uncertainty); the green dot is the S3 position.

The time between S2 and S3 events in experiments [8] was below ~140 μs. Thus, the clouds of unextracted electrons in the RED1 detector were fast-moving - strongly contrasting observations in the LUX detector.
The XeNu detector LLNL is smaller than the RED1 detector (see Fig. 4 for the XeNu design, also [14]). We observed e-burst in the XeNu detector with delays after previous S2 events up to 10-20 ms, so it is not only the large size of the LUX detector which leads to the slowing escape of unextracted electrons from the active surface (under anode).

## 3. Factors preventing surface electron and ion removal

In the LUX detector, all grid holders and field shaping rings are embedded into a PTFE dielectric structure. Only two openings were left in PTFE walls where surface electrons and negative ions can leave the active area and reach grounded electrodes moving along the liquid surface [15].
While no PTFE barriers exist around the liquid surface in the XeNu detector, another effect can impede the escape of surface charges. The fluid level is slightly higher inside the strong electric field region between the extraction grid and the anode (see Fig. 4). Bolozdynya's book on emission detectors [16] mentions this effect. For the extraction voltage of 10 kV applied over about 10 mm gap in between the anode and extraction grid (gate), the liquid Xe level rise is about 0.1 mm. In the XeNu detector's design, we focused on reaching the highest possible extraction electric field. We missed that step on the liquid surface could produce a potential barrier for the surface electron and negative ions to leave the area under the anode. For electrons, moving 0.1 mm step down against the electric field of the order 1kV/mm could be a potential barrier of more than 10 eV (the exact value depends on the distance to the grounded electrodes and other geometry). This potential barrier is greater than temperature, which should lead to unextracted electrons and ions accumulation under the anode.

The anode grid sagging under the electric field's action can also produce a tangential component of the electric field, pushing surface electrons toward the center of the detector. This effect increases with the grid diameter (for fixed wires tension).

The RED1 and ZEPLIN III detectors have no anode grid but a solid aluminum electrode/mirror (see Fig. 2), so anode sagging was effectively absent.

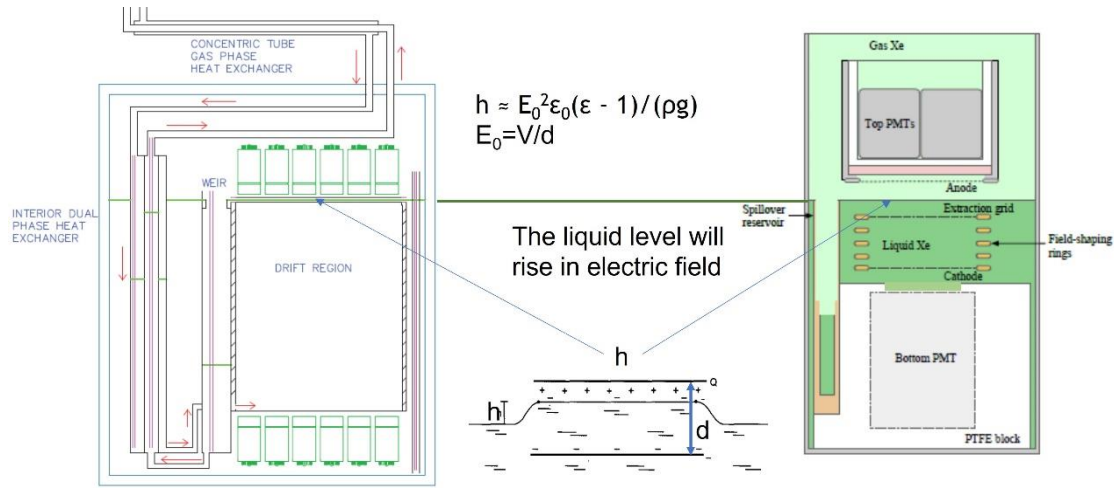

**Figure 4.** Left- part of LUX circulation loop (from [15]) Shown liquid level in the detector, in grounded copper weir reservoir and in dual-phase heat exchangers. Field-shaping ring and grid holders are embedded in PTFE walls (shown as grey hatching), but small openings are left at the liquid level position for liquid to "overflow" into the weir reservoir (to stabilize level position during circulation) and on the opposite side in front of the capacitance level meter (several level meters chowed in magenta). Right- schematic of XeNu detector; there are no PTFE parts around electrodes and grids; liquid level position is stabilized with dielectric weir reservoir, the walls of Xe vessel are stainless steel.
In both detectors there is a small increase of liquid level in a strong electric field applied between extraction grid and anode (dielectric liquid level rise in a capacitor).

The buckling of the extraction grid toward the flat anode in the RED1 and ZEPLIN III detectors produces a tangential field component moving surface electrons out of the center. The RED 100 detector has an anode grid (see [12]). An electrostatic force acting on the anode grid in gas is larger than the force acting on the extraction grid in a dielectric liquid. The asymmetric deformation of the anode grid and extraction grid can lead to surface electrons and ions accumulation.

## 4. Ordered states of surface electrons: Wigner crystallization and dimple lattice

The question arises if some phase transition can occur on the charged liquid Xe (or Ar) surface as more electrons and ions accumulate. Phase transitions and instabilities of a charged liquid surface were extensively studied for electrons and ions on the surface of liquid helium. We need to discuss these helium experiments, though keeping in mind that details of these transitions and instabilities depend on interactions of electrons and ions with the media.

Eugene Wigner predicted the electron ordering phenomenon in 1934. It was observed with electrons localized above the surface of liquid helium at low temperatures in the 1970-s (see for example [17]). It was extensively studied for $^4$He and $^3$He, solid hydrogen, helium and hydrogen films on dielectric substrates, and later for 2-D electron layers in semiconductors heterostructures (in cold 2D electron gas in semiconductors, phenomena like quantum hall effects are present); see the review paper [18].

The appearance of electron ordering strongly affects transport properties. The electric conductivity of the electron layer sitting above helium film on a silicon wafer can drop several orders of magnitude during the Wigner crystallization transition [18]. This conductivity drop happens due to the pinning of the rigid triangular electron lattice to the substrate's local defects and electric field inhomogeneities.

Wigner crystallization requires the repulsion energy of electrons to be larger than the temperature and larger than the energy of the electron's zero-point motion (for a given distance between electrons) [18]. An electric field is required to accumulate a sufficient surface concentration of electrons (of the order of $(3 – 9) \times 10^8$ cm$^{-2}$ in the temperature range 0.40 – 0.65 K – as in the example [17]). At higher temperatures, another instability takes place.

Because of the repulsion of electron and helium atoms at short distances, each electron pressed toward the liquid helium surface by an external electric field forms a microscopic (of atomic scale) dimple. With an increase in the electric field, at some critical field, it became favorable for electrons to form macroscopic dimples with many electrons in each dimple. Surface tension produces repulsion between dimples in addition to Coulomb repulsion and multi-electron dimples from a macroscopic (visible) triangular lattice on the liquid helium surface [19,20]. This lattice can have structural defects like dislocations [20] and can be pinned to inhomogeneities of electric potential and boundaries of the system. The exact size of dimples and the number of electrons in each dimple will depend on the history of charging and field application; for helium, typically, dimples contain about $5 \times 10^6$ electrons and have a depth of a few tenths of a millimeter [20]. Fig.5 showed the results when the surface was initially charged, and then the electric field was increased above the critical value for the transition.

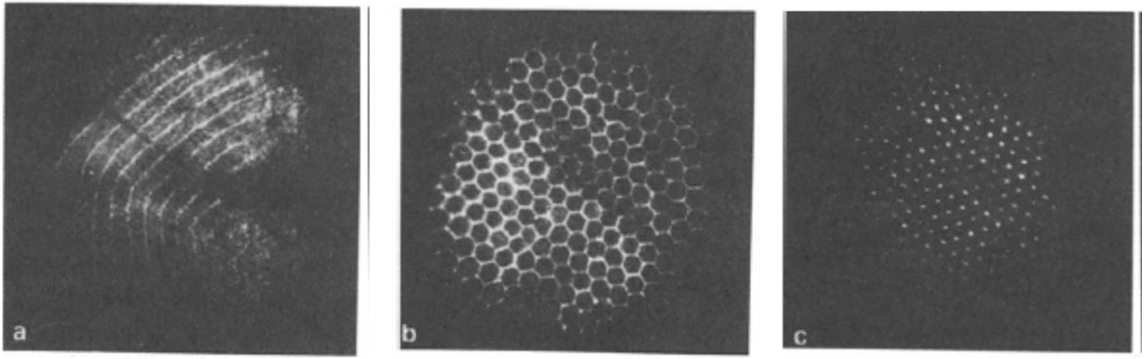

**Figure 5.** Formation of the dimple lattice on a He4 surface (T=3.5K) charged with electrons from above ([19]). Schlieren images (shadowgraph) of the surface deformation approximately 2s (a) and 6s ((b)and (c)) after the field had been increased above critical value E$_c$. The image plane was chosen that convex deformation of the surface (local maxima) appears bright in (a) and (b); in (c) bright correspond to local minima (dimple's bottoms). The distance between adjacent rows of dimples is close to 0.24 cm.

Transition in this scenario usually takes place in two steps. First, wave-like instability with a characteristic wavelength $\lambda_c = 2\pi a$ appears. Then wave-like pattern transforms into a dimple lattice [19,20]. Parameters a and $\lambda_c$ characterize the transition from the capillary to gravity dispersion law for surface waves, and a is called capillary length; $a = \left(\frac{\rho g}{\sigma}\right)^{1/2}$, where ρ is the density of the liquid, σ is liquid surface tension, and g is the gravity acceleration; a is of the order of millimeters, and typical dimple size is about a; see [19,20] for more details.

One also can keep charging liquid surface with electrons from above while keeping the applied electric field larger than the critical value. In this case, dimples will appear individually, one by one, and eventually form a triangular lattice (see [19]).

One can also charge the surface with positive ions from below and get a triangular lattice of hillocks containing multiple ions. Moreover, authors of [19] describe that for some minutes, they were able to stabilize patterns where both positive hillocks and negative dimples were present.

We can summarize that the formation of a dimple lattice (or hillock lattice) behaves as a phase transition and can demonstrate hysteresis [20].

## 5. Geyser emission and electrospraying

With further increase of the electric field, the dimple lattice and hillock lattices on the liquid helium surface also became unstable. The multi-electron dimple on the helium surface can form a bubble with many electrons that "dive" into liquid. Hillocks formed by positive or negative ions can produce charged droplets or jets leaving the liquid. For helium, the effect of droplet formation was named geyser emission and was first observed with flat electrodes parallel to the liquid surface [21]. Recently, this instability was studied in more detail in an inhomogeneous electric field between a 1 mm radius semi-spherical electrode and a flat electrode [22]. This geometry allows producing (stabilizing) one hillock with the size scale of capillary length. Surface deformation and charged droplets/jets formation were filmed for positive and negative ions and for different (tip below or above the liquid surface) electrode orientations (see fig. 6)

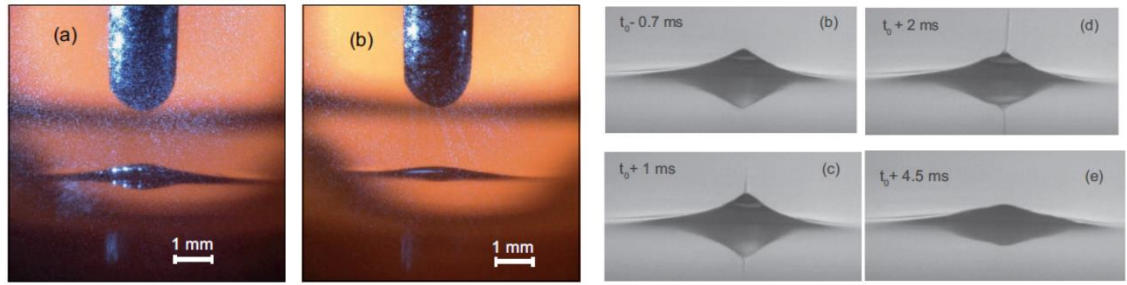

**Fgure 6.** Left color photos- A static deformation of the free surface of superfluid He (Taylor cone) due to the trapped charge in a static electric field; (a) negative charge trapped; (b) positive charge. $T = 2.1$ K.
 Right black and white frames of a fast video recording capturing the process of the charge escape from the Taylor cone. $U_{pin} = -390$ V, $U_{plate} \approx +900$ V (ramp from +800 to +1800 V), $T = 2.1$ K, single-frame exposure time ≈0.19 ms; time $t = t_0$ corresponds to the beginning of the jet emission.

In these experiments, the liquid surface was charged from below with ions and microscopic metal particles produced by laser ablation of a target in the liquid. The electric field ramped up to make the hillock unstable.

No essential difference in electrospraying was reported for these experiments for positively and negatively charged surfaces. The field dependence of emission from liquid helium of electrons (each electron forms a microscopic bubble in liquid helium) and positive helium ions when charged from below are different. There is no onset for electron emission, while a field above about 1.7 keV/cm is required to see for emission (extraction) of positive ions [23]. So, electrons were not present on the surface of liquid helium in the case of the charging from below.

**6. Interpretation of observations in Xe detectors and more questions**

Wigner Crystallization likely can be observed for electrons localized around solid Ar or Xe surface at low temperatures. The other instabilities –formation of waves and hillocks on charged liquid surfaces in an electric field and electrospraying at a high electric field are relatively common for dielectric or slightly conducting liquids. The observations for liquid helium may be closer to other noble fluids. The above discussion cannot provide the complete account or chart boundaries of our transitions and instabilities in detectors. Likely, there are two types of "critical behavior." In one, adding more charges (or an increase in the field) leads to a hillock formation. The other is when the hillock rises uncontrollably (i.e., electrospraying). There could also be changes in single electron emission from the surface and in the extraction of ionization signals. Now we summarize our interpretation of effects in the RED1 detector: as free electrons produced by muon in liquid Xe reach the liquid surface, part of these electrons escape to the gas and produce an S2 electroluminescence signal. A cloud of unextracted electrons moves (linearly in time) toward the detector's perimeter and produces an S3 luminescence signal. S3 Light was detected when the liquid level was low, and the anode support ring was not touching the liquid surface. The drift time of the surface electrons is not exceeding ~140-160 µs. On its way, this cloud emits some electrons into gas. There is a preferential direction for the unextracted electrons to drift, presumably because of the slight tilt of the detector. We can add that a cloud of free electrons can likely move with a surface wave along the surface (high mobility allows the free electron to stay on the wave crest).
Likely, unextracted electrons from ionization events are initially free surface electrons in the LUX detector. When one sees an e-burst, these electrons do not drift from the S2 position. Possibly, charged hillocks are already present on the liquid surface, or many negative charges are present on the surface, so free electrons cannot drift and start to form a new hillock. With enough charges in the hillock, it bursts and injects charged droplets, ions, and electrons into gas. Some droplets can evaporate; negative ions in gas can lose electrons in collisions with neutral atoms.
In the LUX detector, the e-burst was absent on the part of the liquid surface adjacent to the opening in the PTFT wall leading to the copper weir reservoir [4]. At this location, both electrical force and surface flow are helping surface charges to escape, and E-burst instabilities are not happening. As e-burst are absent after significant ionization events here, some mechanism provides sufficiently fast transport of electrons out of the "hot spot" at the S2 position, so an "unstable" hillock is not forming. For the liquid helium surface, electron hopping from hillock to hillock is impossible but could be present on the liquid xenon surface.
In our interpretation, we do not know what is causing delayed single-electron emission events. These events can be free electrons or ions escaping from "stable hillocks" or negative ions coming from the bulk liquid to the surface (with a delay after free electrons) and having some way to escape.
We also do not know how electrons' and ions' surface concentration (density) varies in time after significant events. When it drops to the initial (value before the ionization event), what this initial value is, and what are the mechanisms of motion of the charges along the surface where e-burst are absent?

**7. Effects of surface charges on electron extraction and spectrum of events**

The electric field of surface charges makes the extraction field smaller in the liquid and larger in the gas. The other effect is the inelastic scattering of free electrons on the charged layer. If a "field-overheated electron" loses kinetic energy near the surface, it will stay on the interface. Effects of strong suppression of electron extraction by surface charges were observed in the first measurement of electron extraction efficiency or Ar, Kr, and Xe in [7]. In the experimental cell used in these experiments (see Fig. 7), the side walls are at the same potential as the cathode. For charges trapped at the liquid-gas interphase, reaching a positive electrode (the anode) would be challenging by moving along Xe film on the walls. The paper [7] mentioned the effect of complete suppression of electron emission by liquid surface charging, especially for low voltages applied to the anode. It was possible to restore emissions by reversing the electric field direction for a short time. Unfortunately, we cannot conclude what effect was dominant there; moreover, extraction efficiency in this and any other experiment will be position-dependent if charged hillocks are on the surface.

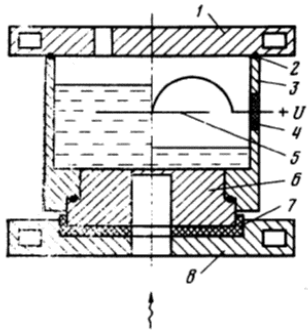

**Fgure 7.** Cell for electron extraction efficiency measurement from condensed noble gases (taken from rom [7]):  1- cover, 2-indium or Teflon gasket, 3- housing, 4- high voltage lead, 5- anode, 6- aluminum cathode liner (radiation from pulse X-ray tube was penetrating into liquid Xe through thin window in this liner), 7- insulating Teflon insert, 8- bottom

The XENONnT experiment presents the most recent example (or puzzle). In this detector, the area under the anode is surrounded by a PTFE wall with no openings for surface charges to leave by drifting along the liquid surface [21]. The effects of grids sagging and the fluid level rise in the applied extraction field should be present in this detector, also leading to electrons and ions accumulation under the anode.

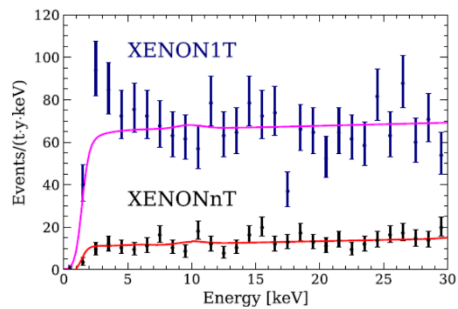

**Fgure 8.** Comparison of event spectra in XENON1T and XENONnT detectors presented by Professor Shingo Kamaza at LIDINE 2023 conference (Warsaw, September 21-23, 2022).

The range of extraction fields in this experiment appears to be limited by an onset of an intense single-electron emission from specific locations on the liquid surface [21], so lower than-designed voltage was applied, and low extraction efficiency should also contribute to unextracted charge accumulation. The number of low-energy events decreases with energy on the XENONnT events spectrum, while for XENON1T, the number of events increases with decreasing energy (see fig.8). Can this be a consequence of electron extraction suppression by the surface charges?

It would be interesting to see the results of "S2 only analysis" for few-electron signals where S1 light is too weak to detect and compare the spectrum of these events in XENONnT to other Xe dark matter detectors. We can ask the same question about LUX results: S2-only analysis was not yet published; paper [4] points to the presence of single electron events and mention that number of few-electron events was insignificant.

**E-bursts and multi-vertex events**

General expectations are that electron extraction can exhibit non-linear effects for significant ionization events. Electrons moving through the liquid surface in a strong extraction electric field should produce local surface overheating at the extraction spot. Also, an electron cloud dragging through the liquid by extraction field transfers momenta to the liquid column, so waves of a jet can be produced on the liquid surface.

We made a striking observation for e-bursts in the XeNu detector: e-bursts were absent after significant single-vertex interactions with particles. The single-vertex event has a symmetric, about one µs duration S2 pulse. For multi-vertex events or muon tracks, the S2 signal is more prolonged, not symmetric, and can have more than one maximum. E-bursts were observed for multi-vertex events with smaller integral S2 intensity than for single-vertex events where e-bursts were absent. Also, the after-luminescence and duration of delayed single-electron emissions were noticeably shorter for single-vertex events. Both local surface heating and jet/wave production can contribute to the observed phenomenon.

As dark matter searches start to consider scenarios where interactions will not be single-vertex-like in the Migdal effect- interactions of surface electrons, surface waves, and shock waves produced in bulk liquid by the primary particle interaction require more attention.

**Discussion and conclusions**

The presence of charges and changes in surface ordering can be detected with two small capacitors by exciting and detecting waves on the charged liquid surfaces [14]. Coupling to surface waves is stronger in the presence of surface charges, and surface ordering changes the wave spectrum (wave propagation velocity). Observation of a single hillock formation and hillock crystal should be possible with optical techniques, like in liquid helium experiments.

One can investigate the effects of a potential barrier at the detector perimeter by placing a ring electrode on the liquid surface around the anode (see Fig. 9). With such an electrode one can controllably change the tangential component of the electric field near the edges. A comparison of events produced by $^{39}$Ar for different potentials applied to the ring electrode can determine if extraction efficiency suppression is present.

One can replace a wire grid anode with a rigid quartz window with an evaporated metal grid or transparent electrodes to avoid anode grid sagging. Larger extraction fields will be beneficial; an increase in the gas pressure above the liquid (i.e., an increase in temperature) can help to increase the extraction field by staying away from gas ionization by electrons around anode wires.

To conclude, statical and dynamic effects associated with free electrons and ions on the surface of noble liquid could be rather complex- as we can see in the example of helium. Uncontrollable accumulation of charges on the liquid-gas interphase in dual-phase Xe and Ar detectors produces uncertainties in data interpretation. It makes difficult direct comparisons of real and parasitic (excess) events spectra in different detectors. It is possible to develop tools for monitoring surface electron density; reducing charge accumulation by changes in the detector design is also feasible. Such changes should minimize uncertainties and allow more reproducible experimentation and studies of mechanisms responsible for the excess backgrounds. Studies of the effects produced by surface electrons and ions will reduce risks associated with the high cost of uncertainties in planned multi-ton liquid Xe and Ar dual-phase detectors.

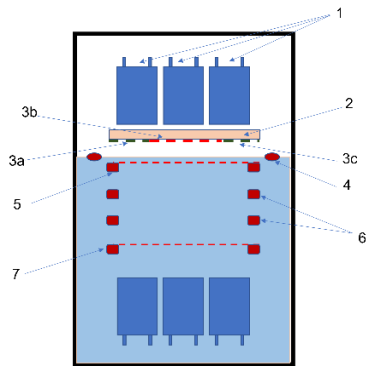

Using ideas and experimental techniques accumulated by helium physics can help work on noble liquid detectors, accelerating searches for physics beyond the standard model.

## Acknowledgments

The author thanks Adam Bernstein for discussions and long-time support of work on the microphysics of dark matter and CEvNS detectors; Professor Paul Leiderer for introduction to the surface electrons physics that the author got in Paul Leiderer's lab back in 1993, and insightful discussions during writing of this paper; Dr. Teel Pershing and Eli Mizrahi for help with running the XeNu detector to allow important observations; outside of the scope of LLNL noble liquids group projects; Dr. Jingke Xu and Dr, Dmitry Akimov for discussions and explaining results from the LUX and RED1 detectors. This work was performed under the auspices of the US Department of Energy by Lawrence Livermore National Laboratory under Contract DE-AC52-07NA27344. The author acknowledges LDRD grant 20-SI- 003 and DOE fieldwork proposal number SCW1508. LLNL-CONF-842379.

## References

[1] P. Sorensen et al., "Lowering the low-energy threshold of xenon detectors," Proc. Sci., IDM20102011 (2011) 017[arXiv:1011.6439].

[2] D.S. Akerib et al., "First Dark Matter Search Results from the LUX-ZEPLIN (LZ) Experiment," https://doi.org/10.48550/arXiv.2207.03764

[3] J. Aalbers et al., "A Next-Generation Liquid Xenon Observatory for Dark Matter and Neutrino," *J. Phys. G: Nucl. Part. Phys.* **50** 013001(2023), **DOI** 10.1088/1361-6471/ac841a

[4] D. S. Akerib et al., An investigation of the background electron emission in the LUX detector, Phys. Rev. D 102, 092004 (2020).

[5] S. Pereverzev, "Detecting low-energy interactions and the effects of energy accumulations in materials," Phys. Rev. D 105, 063002 (2022),

[6] P. Adari et al., Excess workshop: Description of rising low-energy spectra," https://doi.org/10.21468/SciPostPhysProc.9.001

[7] E. M. Gushchin, A. A. Kruglov, V. V. Litskevich, A. N. Lebedev, I. M. Obodovski, and S. V. Somov, "Electron emission from condensed noble gases." Zh. Ebp. Teor. Fiz. 76, 1685-1689 (1979)

[8] E.M. Gushschin, A. A. Kruglov, and I.M. Obodovkii, "Emission of "hot" elections from liquid and solid argon and xenon," Sov. Phys. JETP 55, 860 (1982), http://www.jetp.ras.ru/cgi-bin/dn/e_055_05_0860.pdf.

[9] A. G. Khrapak, W.F Schmidt and E Illenberger, "Charged Particle in Bulk and Near the Surfce of a Non-Polar Liquid Dielectric," *Proceedings of 2002 IEEE 14th International Conference on Dielectric Liquids. ICDL 2002 (Cat. No.02CH37319)*, Graz, Austria, 2002, pp. 71-77, **DOI:** 10.1109/ICDL.2002.1022697

[10] D.Y. Akimov et al., Observation of delayed electron emission in a two-phase liquid xenon detector. JINST 11, C03007 (2016).

[11] D.Y. Akimov et al., "The ZEPLIN-III dark matter detector: Instrument design, manufacture and commissioning," Astroparticle Physics, V. 27, pp.46-60, (2007).

[12] D. Akimov et al., "First ground-level laboratory test of the two-phase xenon emission detector RED-100" JINST **15,** P02020 (2020).    DOI 10.1088/1748-0221/15/02/P02020

[13] D. Akimov, private communication

[14] J. Xu, S. Pereverzev, B. Lenardo, J. Kingston, D. Naim, A. Bernstein, K. Kazkaz, and M. Tripathi, Electron extraction efficiency study for dual-phase xenon dark matter detector", Phys. Rev. D 99, 202 103024 (2019).

[15] A.W. Bradley, D. S. Akerib, et al., "LUX Cryogenics and Circulation," Physics Procedia 37 ( 2012 ) 1122 – 1130.

[16] Alexander Bolozdyna, "Emission Detectors," World Scientific Publishing Company; 1st edition (July 30, 2010).

[17] C.C. Grimes and G. Adams, "Evidence of a Liquid-to-Crystal Phase transition in a Classical, Two-Dimensional Sheet of Electrons," Phys. Rev. Lett., 42, pp.795-798 (1979).

[18] Monarkha, Yu P., and V. E. Syvokon. "A two-dimensional Wigner crystal." *Low-Temperature Physics* 38.12 (2012): 1067-1095.  https://doi.org/10.1063/1.4770504

[19] P. Leiderer, W. Ebner, V.B. Shikin, "Macroscopic electron dimples on the surface of liquid helium," Surface science 113, pp. 405-411 (1982).

[20] P. Leiderer, "Electrons at the surface of quantum systems," *J Low Temp Phys* **87**, 247–278 (1992). https://doi.org/10.1007/BF00114906

[21] Volodin, A. P., and M. S. Khaikin. "Ion"geysers"on the surface of superfluid helium." *JETP Lett. (USSR)(Engl. Transl.);(United States)* 30.9 (1979).

[22] P. Moroshkin, P. Leiderer, Th. B. Möller, and K. Kono "Taylor cone and electrospraying at a free surface of superfluid helium charged from below," PHYSICAL REVIEW E **95**, 053110 (2017).

[23] P.P. Boyle and A.J. Dahm, "Extraction of Charged Droplets from Charged Surfaces of Liquid Dielectrics," Journal of Low Temperature Physics, 23, pp.477-486 (1976). https://doi.org/10.1007/BF00116935